\documentclass[aps,prl,twocolumn,showpacs,preprintnumbers]{revtex4}
\usepackage{graphicx}
\usepackage{bm}

\newcommand{\be}[1]{\begin{equation} \label{(#1)}}
\newcommand{\ee}{\end{equation}}
\newcommand{\ba}[1]{\begin{eqnarray} \label{(#1)}}
\newcommand{\ea}{\end{eqnarray}}
\newcommand{\vev}[1]{{\langle {#1} \rangle }}

\def\bi{\bibitem}

\begin{document}

\preprint{\sf Version final (\today)}
\title{A model of CP Violation from Extra Dimension}
\author{Darwin Chang$^{1,2}$, Chian-Shu Chen$^1$, Chung-Hsien Chou$^3$, Hisaki Hatanaka$^1$ }
\altaffiliation[]{Authors listed in alphabetic order.}
\affiliation{$^1$ Physics Department, National Tsing-Hua
University, Hsinchu 300, Taiwan, ROC}
\affiliation{$^2$ National
Center for Theoretical Sciences, P.O. Box 2-131,  Hsinchu, 300,
Taiwan, ROC}
 \affiliation{$^3$Institute of Physics, Academia
Sinica, Taipei 115, Taiwan, ROC}

\date{\today}

\begin{abstract}
We construct a realistic model of CP violation in which CP
is broken in the process of dimensional reduction and orbifold
compactification from a five dimensional theories with $SU(3)\times SU(3) \times SU(3)$ gauge symmetry.  CP violation is a result of the Hosotani type gauge configuration in the higher dimension.
\end{abstract}

 \maketitle

\section*{Introduction}

There are many mechanisms of obtaining CP violation if one starts from a CP conserving higher dimensional theory\cite{thirring,cpx1}.  The idea is not new.  For example, Thirring considered such possibility as early as 1972\cite{thirring}.
Recently, with the renewed interest in the extra dimensional theories due to many new approaches to the additional dimensions, new schemes of obtaining CP violation from a CP conserving higher dimensional theory were proposed.  More recently**, for example, in Ref.\cite{CM}, CP violation arises as a result of compactification due to the
incompatibility between the orbifold projection condition that defines the projected geometry of the space and the higher
dimensional CP symmetry.  Therefore, the origin of CP violation can be geometrical in nature.

Another interesting geometrical scheme has been pursued in Ref.\cite{frere}.  In this scheme, CP violation arises out of the possibility that the high dimensional gauge
field may develop a nontrivial configuration when compactified on an orbifold
type of geometry with nontrivial topological loops.  Using such configuration to break gauge symmetry is called the Hosotani mechanism \cite{hoso}.  It was initially proposed
to break gauge symmetry, however, it was realized \cite{frere} that it can also
be used to break CP symmetry.  In Ref.\cite{frere2}, several interesting
 models were pursued in this direction.  They include one with
 $SU(2)_L\times SU(2)_R \times U(1)_{B-L}$ gauge symmetry and another** one
   with $SU(4)\times U(1)$ gauge symmetry in the higher dimensional theory.
Some prospects of grand unification in the high dimension were also discussed.  The models aim to produce the Kobayashi-Maskawa Model in four dimension.

One weakness in the models proposed in Ref.\cite{frere,frere2} is that
 the Hosotani vacuum expectation value are used to break the electroweak $SU(2)_L$ symmetry.
   Since the Hosotani vacuum expectation value is related to the ground state
 value of the Wilson loop integral over the compactified dimension, as we shall illustrate in the next section, it is expected that its value should be of the size of inverse of the
     extra dimension, $R^{-1}$.  Since the current experimental limit
      on $R^{-1}$ is larger than the weak scale already, it is preferable
      to have a model in which the Hosotani mechanism while breaks CP and
       gauge symmetry but does not involve in $SU(2)_L$ electroweak breaking.

In this paper, we propose a higher dimensional model in which the Hosotani
mechanism breaks CP and higher dimensional gauge symmetry.  The electroweak
 gauge symmetry is broken in four dimension by a zero mode which
  is $SU(2)_L$ scalar doublet.  To achieve this goal, we employ the
trinification gauge group of $SU(3)^3$ in higher dimension
  in which the extra fermions are naturally needed.
   In particular, in the quark sector, the new fermions
   are the extra vectorial down type quarks, $D_{L,R}$.
    The Hosotani mechanism induces CP violating mixing
     between the light quark and the heavy down quark as
     well as breaking the $SU(3)^3$ gauge symmetry.
      Such mixing results in four dimensional Kobayashi-Maskawa CP violation among the light quarks.

At the bottom of the issue, the CP violation arises because the
 Hosotani vacuum expectation, being related to the Wilson line
  of the gauge configuration, is pseudo-scalar and effectively
  CP-odd in nature in four dimension.  In some simple cases,
  such CP violating phase can be rotated away, but in general it can not.

\section{Hosotani breaking and its scale}
If the space is not simply connected, the gauge field can develop a
vacuum expectation value along a non-contractable loop in the extra
dimension and it cannot be gauged away\cite{hoso}.  The VEV of this gauge field
can give rise to a mass term for fermions (called ``Hosotani mass term''):
\begin{eqnarray}
 -ig \bar{\psi} \langle A_y \rangle \psi.
\label{eq:massterm}
\end{eqnarray}

It is not hard to show that the magnitude of the Hosotani mass term should** be, in
general, of the order of the compactification scale.  For example, consider** a
$SU(N)$ gauge theory on a five dimensional non-simply connected space
$M^4 \times S^1$, where $M^4$ is Minkowski space and the radius of circle $S^1$ is $R$,
with $N_f$ flavor fermions in fundamental representation.   The** 1-loop effective potential takes the form \cite{hoso,gauge-higgs}:
\begin{eqnarray}
 V_\mathrm{eff}(\langle A_y \rangle) &=& 
   \frac{3}{128\pi^7R^5}
 \left\{ -3\sum_{i,j=1}^{N} F_5(\theta_i-\theta_j)
 \right. \nonumber \\ 
 && +  \left. 2^2 N_f \sum_{i=1}^{N} F_5(\theta_i)  \right\} ,
\label{eq:effpot}
\end{eqnarray}
where $ F_5(x)  = \sum_{n=1}^{\infty} \cos (nx)/ n^5 $
and gauge vacuum expectation values are parameterized as $\langle A_y
\rangle = (2\pi g_5R)^{-1} \mbox{diag} (\theta_1, \theta_2, \cdots, \theta_N)$.
In the r.h.s of (\ref{eq:effpot}),  the first and second terms are
the gauge-ghost and fermion 1-loop contributions**, respectively.
Since the $F_5(x)$ is a cosine-like function with period $2\pi$, the nonzero**
minimum of (\ref{eq:effpot}) tends** to have a minimum at $\theta_i \sim
{\cal O}(1) $ (or $\langle A_y
\rangle \sim R^{-1}$) unless there are extra fine-tunings.
%  Furthermore, since the masses of $A_y$ is
%regarded as a Hessian of the effective potential $ (gL)^2 \cdot
%\partial^2 V_\mathrm{eff}/\partial \theta_i \partial \theta_j$, they are
%also closely related to the compactification scale

Therefore, when we consider models with compactification radius
much smaller than electroweak scale, it is unnatural to use Hosotani mechanism to break
$SU(2)_L$ gauge symmetry.

%some see-saw like mechanism will be
%needed to give light masses to standard Model fermions.

\section{The $SU(3)_c\times SU(3)_l\times SU(3)_r$  model}

In this paper we propose a $SU(3)_c\times SU(3)_l\times SU(3)_r$ gauge
theory\cite{tri}  which is assumed to be CP symmetric in $4+1$ dimension.
Orbifold symmetry breaking mechanism breaks the gauge symmetry
  to $SU(2)_L\times SU(2)_R\times U(1)_L\times U(1)_R$ when the
  space is compactified on orbifold to a 3+1 dimension.
  The zero modes of the compactification, serving as the four dimensional scalar bosons, further break the symmetry to Standard Model group and then to $U(1)_{em}$.

%\paragraph{Particle content}
Let's first list the basic field contents of this model in $4+1$ dimension:\\
\begin{center}
\begin{tabular}{|c||c|}
\hline  & $SU(3)_c \times SU(3)_l \times SU(3)_r $\\
\hline
\hline  $A_{c,M}$&  $(8,1,1)$\\
\hline  $A_{l,M}$&  $(1,8,1)$\\
\hline  $A_{r,M}$&  $(1,1,8)$\\
\hline $Q_{l} $&  $(3,3,1)$\\
\hline  $\bar{Q}_{r}$&  $(\bar{3},1,\bar{3})$\\
\hline  $L$ &  $(1,\bar{3},3)$\\
\hline  $\Phi_{l}$&  $(1,8,1)$\\
\hline  $\Phi_{r}$&  $(1,1,8)$\\
\hline  $\Sigma_{\bar{l}r}$&  $(1,\bar{3},3)$\\
\hline
\end{tabular}
\end{center}
where the index $M=(\mu,y)$ runs from 0 to 4, $\mu$ from 0 to 3
and $y$ is the fourth dimension.  Note that an irreducible fermion
in $4+1$ dimension contains fermions of both chirality in $3+1$
dimension. In this trinification model, the gauge fields
$A_{c}$,$A_{l}$, and $A_{r}$ are in the adjoint representation of
their perspective $SU(3)$.  The fermionic fields $Q_{l}
(\bar{Q}_{r})$ contain the standard model
left-handed (right-handed) quarks and their chiral partners as required in a $4+1$ dimensional theory. The lepton multiplet, $L$, contains the leptonic sector of
the standard model and additional leptons (to be discussed later) as well as their chiral partners.  The scalar fields $\Phi_{l}$, $\Phi_{r}$ and
$\Sigma_{\bar{l}r}$ are needed to give masses to particles.  One
also notes that this model can naturally be embedded into a grand
unified group, $E_6$, if so desired.

To break gauge symmetry by geometry through the Hosotani mechanism,
we compactify the $4+1$ dimensional space on orbifold.
Orbifold is produced by imposing projection condition on
the space and the fields.
This projective symmetry dictates a transformation on each
field and selects the zero modes which will serve as the low
energy modes that play the active role in $3+1$ dimensional theory.
   In this paper, we consider the simplest case in which there
    is only one extra dimension with a $Z_2$ projection, i.e.,
    $S^1/Z_2$.  It is the circle with the points identification
    under the parity operation in the fourth dimension ( $y \rightarrow -y$).

Now we have to specify the $Z_2$ representation of each field.
  Note that, for any transformation under $Z_2$, we are allowed
  to insert the transformation matrix which belongs to
   symmetry of the theory, such as a discrete gauge transformation
   $P_{\mathcal{G}} \in \mathcal{G}$ (with $P_{\mathcal{G}}^2 = \mathbf{I} $).

So we have boundary conditions
\begin{eqnarray}
 A_\mu^a (x^\nu,y)\; \lambda^a
& = & A_\mu^a (x^\nu,-y) \; P_{\mathcal{G}} \lambda^a
P_{\mathcal{G}}^{-1},
\nonumber\\
 A_y^a (x^\nu,y) \;\lambda^a
&= &
 - A_y^a (x^\nu,-y) \; P_{\mathcal{G}} \lambda^a P_{\mathcal{G}}^{-1},
 \nonumber
\end{eqnarray}
for gauge fields and
\begin{eqnarray}
\Psi(x^\mu,y) = P_{\mathcal{G}} \gamma_5 \Psi(x^\mu,-y),
\nonumber\end{eqnarray} for fermions \cite{note1}.
%
%\footnote{With phases, $\Psi(x_\mu,y) = \eta P_{\mathcal{G}} \gamma_5
%\Psi(x_\mu,-y)$, Here we have a put a phase factor $\eta, (| \eta |^2 =
%1)$ in the fermion transformation. For different fermions $\Psi_i$, the
%phase factor $\eta_i$ could be different and the relative phases between
%fermions could have nontrivial effects.}

The transformation properties for scalars are determined by their
couplings to fermions.  Since, under the $Z_2$ transformation,
 $\bar{\Psi}_i\Psi_j$ term transforms into
%
%\footnote{With phases, $- \eta_i^* \eta_j\bar{\Psi}_i\Psi_j$}
%
 $ - \bar{\Psi}_i\Psi_j$, the scalar fields must transform as
\begin{eqnarray}
 \Phi^a (x^\mu,y) \; \lambda^a = - \Phi^a
(x^\mu,-y) \; P_{\mathcal{G}} \lambda^a P_{\mathcal{G}}^{-1},
\nonumber
\end{eqnarray}
for the scalar boson to couple to the fermion.  An adjoint scalar $\Phi$,
 which couples to fermions, and the fourth component of the gauge
  field $A_y$ must gets the same zero modes after compactification.
%
%\footnote{Note that if $\eta_i^* \eta_j = -1$,
%the scalar must be even under the $Z_2$ parity.}
%
To illustrate the orbifold projection, let's first consider the
case of only one $SU(3)$. The representation decomposes as
follows:
\begin{center}
\begin{tabular}{ccl}
\hline
$ SU(3)$             & $\supset$    & $SU(2)  \times U(1)$ \\
\hline\hline
 $3$    & $\rightarrow$ & $2_{1}+1_{-2} $\\
 $8$    & $\rightarrow$ & $3_{0}+1_{0}+2_{3}+2_{-3}$  \\
\hline
\end{tabular}
\end{center}
here we have used $P_{\mathcal{G}} = \mbox{diag}(-1,-1,1)$ as the appropriate
projection.  We verify easily that $P_{\mathcal{G}}$ commutes
with generators of an $SU(2) \times U(1)$ subgroup, while anticommuting
with the other generators, say $ [ \lambda_{1,2,3,8}\, , P_{\mathcal{G}}
] = 0 $, $ \{ \lambda_{4,5,6,7}\, , P_{\mathcal{G}} \} = 0 $.
As a result, the zero mode gauge fields are:
\begin{eqnarray}
A_\mu^{\bar{a},(0)} & \rightarrow &1_{0}+3_{0}, \quad (\bar{a}= 1,2,3,8)
 \nonumber \\
A_y^{\hat{a},(0)} & \rightarrow & 2_{3}+2_{-3}, \quad (\hat{a}=
4,5,6,7). \nonumber
\end{eqnarray}
The fermions in the $3$ representation reduce to the following zero
modes
\begin{equation}
\left( u^{(0)}_{L} \, d^{(0)}_{L} \, B^{(0)}_{R} \right)
 \leftrightarrow 2_{1} \oplus 1_{-2}.
\nonumber
\end{equation}
Here and from now on, $L, R$ represent the chirality of the $3+1$ dimensional fermions..
Zero modes of adjoint scalar $\Phi$ coupled to fermions have the same gauge quantum numbers as $A_y$.

More precisely, since $(3,3,1) \mapsto (3,2,1) \oplus (3,1,1)$, we
can write, for example, $Q_{l} $ as :
\begin{eqnarray}
 Q_{l} = \left(%
\begin{array}{c}
  u_l^{\alpha} \\
  d_l^{\alpha} \\
  B_l^{\alpha} \\
\end{array}%
\right) \Rightarrow_{orbifolding} \left(%
\begin{array}{c}
  u_{lL}^{\alpha,(0)} \\
  d_{lL}^{\alpha,(0)} \\
\end{array}%
\right) \oplus B_{lR}^{\alpha,(0)},
\nonumber
\end{eqnarray}
where $\alpha = 1,2,3 $ is the $SU(3)_c$ group index,
the superscript $(0)$ denotes the zero modes in $3+1$ dimensions.  Similar notations
can be used to $\bar{Q}_{r} $
 field. Note that here we use $l,r$ to denote gauge groups while using
 $L,R$ to denote the handedness of the fermions.
>From now on, we neglect the color index $\alpha $ and zero mode label $(0)$.

Back to trinification model, in order to break $SU(3)_l \times SU(3)_r$ down to $SU(2)_l \times
SU(2)_r \times U(1)_l \times U(1)_r$ by orbifolding,
we choose our projection operator $P_{\mathcal{G}}$ as
\begin{equation}
P_{\mathcal{G}} = \mbox{diag}(-1,-1,1)_l \otimes
\mbox{diag}(1,1,-1)_r.
\end{equation}
After the projection, the field content of the zero modes of the theory
in $3+1$ dimension becomes:\cite{note2}
%
%\footnote{We assume that the $Z_2$ transformation property
%between $Q_{l}$ and $\bar{Q}_r$ has a relative phase such that $\eta_R^*
%\eta_L = -1$.}
%
\noindent
\begin{center}
\begin{tabular}{|c|c||c|}
\hline
   & $ SU(3)_l \times SU(3)_r $ & $SU(2)_l \times SU(2)_r $ \\
   & & $ \times U(1)_l \times U(1)_r $ \\
\hline
\hline $A_{l\mu}$ & $(8,1)$ & $(3,1)_{(0,0)}+(1,1)_{(0,0)}$\\
      $A_{ly} $&        &  $(2,1)_{(3,0)}+(2,1)_{(-3,0)}$\\
\hline $A_{r\mu}$ & $(1,8)$ & $(1,3)_{(0,0)}+(1,1)_{(0,0)}$\\
      $A_{ry} $&  &  $(1,2)_{(0,3)}+(1,2)_{(0,-3)}$\\
\hline $\left(%
\begin{array}{c}  u_{l} \\ d_{l} \\ \end{array}  \right)_L$
      & $(3,1)$ & $(2,1)_{(1,0)}$\\
      $B_{lR} $&      &  $(1,1)_{(-2,0)}$\\
\hline $\left(%
\begin{array}{c}   u^c_{r} \\ d^c_{r} \\ \end{array} \right)_R$
 & $(1,\bar{3})$ & $(1,\bar{2})_{(0,-1)}$\\
      $B^c_{rL} $&            &  $(1,1)_{(0,2)}$\\
\hline $L_L$ & $(\bar{3},3)$ & $(\bar{2},2)_{(-1,1)}+(1,1)_{(2,-2)}$\\
      $L_R$&   &  $(\bar{2},1)_{(-1,-2)} + (1,2)_{(2,1)}$\\
\hline $ \phi_l$ & $(8,1) $&  $(2,1)_{(3,0)}+(2,1)_{(-3,0)}$\\
\hline $ \phi_r$ & $(1,8) $&  $(1,2)_{(0,3)}+(1,2)_{(0,-3)}$\\
\hline $\sigma_{\bar{l}r}$ & $(\bar{3},3)$ & $(\bar{2},2)_{(-1,1)}$\\
      $\chi $&   &  $(1,1)_{(2,-2)}$\\
\hline
\end{tabular}
\end{center}
where the $SU(3)_l \times SU(3)_r$ origins of the fields are also included in the middle column.
Note that $B_{lR} $ and $B^c_{rL} $ form a vector-like quark pair singlet under $SU(2)_L$.
We can identify $B_{lR} $ and $B^c_{rL}$ as heavy quarks
$D_{L,R}$.

Because the scalar fields $\Phi_l, \Phi_r$ and $\Sigma_{\bar{l}r}$
transform differently under $Z_2$, their respective zero mode fields in $3+1$
dimension after the orbifolding also have different transformation property.\\

\section{CP violation}

In order to make a mass term for quarks, some scalar field must
develop VEVs and break $SU(2)_l \times SU(2)_r$.  In this model, we assume that $\phi_r$ and $\chi$
develop VEVs. $\vev{\phi_r}$ and $\vev{\chi}$ break the
$SU(3)_c\times SU(2)^2 \times U(1)^2$ to $SU(3)_c\times SU(2)_L
\times U(1)_Y$.
In order to get $M_D$ (mass term for $\bar{B}_{lR} B^c_{rL}$), we
need a standard model singlet field $\chi$ to survive after
orbifolding. That's why we need $\Sigma_{\bar{l}r}$ field. Note
that the VEV of $\chi$ breaks $U(1)_l \times U(1)_r$ to $U(1)_Y$ and gives
$M_D \sim f_{\Sigma} \vev{\chi}$.
In order to couple $\bar{d}_r$ with $B_{rL}$, we need the
$SU(3)_{r}$ scalar field $\phi_{r}$ to develop VEV along the
$\lambda_6$ direction, say $\vev{\phi_{r}}= \vev{\phi_{r6}}=v_r \lambda_6$.
$\vev{\phi_r}$ is related to the $SU(2)_r$ breaking scale.
In that case, the Hosotani mass term $\vev{A_{ry}}$ which is assumed to develop a
nonzero VEV here, is then forced\cite{note3}
to be parallel to $\vev{\phi_r}$, that is, $\vev{A_{ry}} = v_A \lambda_6$,
by the minimization condition of the Hosotani potential\cite{gauge-higgs}.
Note that both  $\vev{\phi_r}$ and $\vev{A_{ry}}$ are of the order $R^{-1}$.

As a result, for the down sector, we have the following mass matrix:
\begin{equation}
\pmatrix{ \bar{d}_{rR} & \bar{B}_{lR} }
\left( \begin{array}{cc}
 \hat{f}_{\Sigma} v_{\sigma}   &  \hat{f}_r v_r  + i g_r v_A \hat{\bf 1}
\\
 \hat{f}_l v_l  & \hat{M}_D\sim \hat{f}_\Sigma \vev{\chi}
\end{array} \right)
\pmatrix{  d_{lL} \cr  B_{rL} },
\label{eq:mass}
\end{equation}
where $\hat{f}_l$, $\hat{f}_r$, $\hat{f}_\Sigma$  denote the
Yukawa coupling matrices of $ \phi_l $, $ \phi_r $, $\Sigma$,
respectively.  Note that the Hosotani term is generation independent and proportional to the unit matrix $\hat{\bf 1}$.  Generically, this fermion mass matrix will give complex phases and
lead to at least Kobayashi-Maskawa(KM) type of CP violation.
Note that the up quark mass matrix is purely
from $\hat{f}_\Sigma v_{\sigma}$ which is real and does not contribute to the CP violating phase.

The down quark mass matrix above have some similarity with that
in a model proposed to solve the strong CP problem \cite{ck}.
Unfortunately the current model as it is does not provide a
solution to strong CP problem.   However, an extension of
the model with flavor symmetry may be able to achieve this goal.

Note that we assumed that $\vev{ A_l }$ vanishes  because we don't
want to use Hosotani term to break $SU(2)_l$ gauge symmetry. We
want to break the $SU(2)_l$ group at a scale lower than the compactification
scale through the Higgs mechanism as in the Standard Model.\\

\section{neutrino mass}

In the lepton sector, $\vev{\phi_r}$ will give large masses to two
of the three $SU(3)_l$ triplets and leave one triplet per
generation in $L$ at the usual $SU(2)_l$ scale which serve as the usual
light leptons.  The $v_{\sigma}$ will give rise to Dirac masses for
the lepton.  The model as it is still have massless neutrinos.

Let's consider the leptonic sector in more detail. We fix the
convention such that $L \rightarrow U_l^+ L U_r, \bar{L}
\rightarrow U_r^+ \bar{L} U_l$, and $\Phi_r \rightarrow U_r^+
\Phi_r U_r$.

The mass term of the leptons are coming from three types of terms:
$Tr(\bar{L}L\Phi_r), Tr(\bar{L}\Phi_l L)$ and $LL\Sigma$
\footnote{ Write explicitly, this is $\epsilon_{lmn}\epsilon^{ijk}
L^l_i L^m_j \Sigma^n_k$.}.

If we write
\begin{equation}
L \sim \left(%
\begin{array}{ccc}
  N_L^0 & E_{2L}^+ & E_{2R}^+ \\
  E_{1L}^- & N_{1L} & N_{1R} \\
  E_{1R}^- & N_{2R} & N_{2L} \\
\end{array}%
\right)
\end{equation}
and
\begin{equation}
\Sigma \sim \left(%
\begin{array}{ccc}
  S_{11}^0 & S_{12}^+ & 0 \\
  S_{21}^- & S_{22}^0 & 0 \\
  0 & 0 & \chi \\
\end{array}%
\right),
\end{equation}
we get the following interacting terms from $LL\Sigma$:
\begin{eqnarray}
LL\Sigma \supset && N_{1L} N_L^0 \chi - E_{1L}^- E_{2L}^+ \chi
\\ \nonumber &+& N_{1L} N_{2L} S_{11}^0 - N_{1R} N_{2R} S_{11}^0
\\ \nonumber &+& N_L^0 N_{2L} S_{22}^0 - E_{1R}^- E_{2R}^+
S_{22}^0 \\ \nonumber &+& N_{2R} E_{2R}^+ S_{21}^- - N_{2L}
E_{2L^+} S_{21}^- \\ \nonumber &+& N_{1R} E_{1R}^- S_{12}^+ -
N_{2L} E_{1L}^- S_{12}^+.
\end{eqnarray}

Combining with terms from $Tr(\bar{L}L\Phi_r), Tr(\bar{L}\Phi_l L)$,
we get the following charged lepton mass matrix, in the basis of
$( E_{1R}^- , \bar{E_{2L}^+} , E_{1L}^- , \bar{E_{2R}^+} )$,
\begin{eqnarray}
M_{charged } \sim \left(%
\begin{array}{cccc}
  0 & 0 & <\phi_{l6}> & <S_{22}^0> \\
  0 & 0 & <\chi> & <\hat{\phi}_{r6}> \\
  <\phi_{l6}> & <\chi> & 0 & 0 \\
  <S_{22}^0> & <\hat{\phi}_{r6}> & 0 & 0 \\
\end{array}%
\right)
\end{eqnarray}
while the neutral lepton mass matrix, in the basis
of $(N_L^0 , \bar{N_{1R}} , \bar{N_{2R}}  , N_{2L} , N_{1L} )$, is of the form,
\begin{eqnarray}
& M_{neutral } \sim & \nonumber\\
&\left(%
\begin{array}{ccccc}
  0 & 0 & 0 & <S_{22}^0> & <\chi> \\
  0 & 0 & <S_{11}^0> &  <\phi_{l6}> & <\hat{\phi}_{r6}> \\
  0 & <S_{11}^0> & 0 & <\hat{\phi}_{r6}> &  <\phi_{l6}> \\
  <S_{22}^0> &  <\phi_{l6}> & <\hat{\phi}_{r6}> & 0 & <S_{11}^0> \\
  <\chi> & <\hat{\phi}_{r6}> &  <\phi_{l6}> & <S_{11}^0> & 0 \\
\end{array}%
\right)&
\end{eqnarray}
where $<\hat{\phi}_{r6}> $ is a linear combination of
$<\phi_{r6}> $ and $i <A_{ry}> $ and give rise to CP violation in the lepton sector.
The Yukawa couplings in these matrices are ignored since they are meant to indicate only the scale of the respective terms.
Note that $<S_{11}^0>,  <S_{22}^0>$ and $ <\phi_{l6}>$ are $SU(2)_l$ breaking scale and should be  small compared with $<\chi>$ or $ <\phi_{r6}>$ which are $SU(2)_r$ breaking scale or higher.  To first approximation, we can set $SU(2)_l$ breaking scale to zero and we find that there are two
zero eigenvalues in the charged mass matrix and only one zero
eigenvalue in the neutral one.  That means that we have one vectorial pair of
massless charged leptons and one massless neutral chiral lepton. Turning on
$<S_{11}^0>, <S_{22}^0>$ and $<\phi_{l6}>$ will make the determinants
nonzero and we naturally have a see-saw structure in the neutral lepton mass
matrix. That means that we have a pair of light charged leptons
and a super-light neutral lepton (we identify it as the neutrino per generation) due
to see-saw mechanism.

\section{Discussions}

The problem of using Hosotani mechanism to break gauge symmetry
and generate masses is that the natural value of the generated mass scale is of order
of compactification scale. This is because the Hosotani mechanism makes use of
the phase factor of the Wilson line integral $\exp {(i g_r \int A_{ry} dy)}$
which in term gives $g_r \langle A_r \rangle \sim O(\frac{1}{R})$,
where $g_r$ is the gauge coupling constant of $SU(2)_r$ and $R$ is
the radius of the extra dimension.

To decouple the Hosotani breaking of CP symmetry with the lower energy
$SU(2)_l$ breaking, we have constructed a realistic model which can generate
CP violations even though all the Yukawa couplings and all the Higgs VEVs are real.
Furthermore, we can have see-saw type neutral lepton mass matrix naturally.

%For conventional trinification model, people usually assume
%$g_c = g_l = g_r$, in our model we don't need to require this.

One important feature of our model is that we use Hosotani
term $g_r \langle A_r \rangle$ to break $SU(2)_r$ gauge symmetry
and to** generate CP violation at the same time!  The breaking of gauge symmetry
and that of CP are related to each other and they are both originated from the existence of extra
dimensions.  In addition, by assigning proper gauge and $Z_2$ quantum numbers
for each fields in the $4+1$ dimensional theory, we have a natural way to provide
 the desired chiral state in the $3+1$ dimensions.

So, we manage to account for a $3+1$ dimensional theory which is effectively
KM-like in nature starting from a higher dimensional CP conserving theory
using gauge as well as CP breaking Hosotani mechanism.  This may provide some
insight to the origin of CP violation.  A few questions immediately arise.  Is strong CP problem of the
KM model resolved in this model?  The question is unfortunately no.  However, the mechanism used here seems to provide enough flexibility that one suspects that there may be similar models properly extended that can solve the strong CP problem.  Another question is the new physics that may arise related to this mechanism.  The new physics scale $R^{-1}$ can in principle be very high such as $10^{10-11}$ GeV if one uses the see-saw mechanism to explain the small observed neutrino masses.  In that case, the CP violating scale, the compactification scale and the neutrino see-saw scale are all tie together which is interesting.
On the other hand, if one first ignores the lepton sector, since it is more remote to the observed CP phenomena in the quark sector, or if one allows fine-tuning to make the neutrino masses small, then the scale $R^{-1}$ can be as small as the experimental limit on the light vectorial quark mass.  Since such vectorial quark can in principle violate the unitarity of the quark mixing matrix, one can derive a limit of around a few TeV based on the experimental limit on the unitarity of KM mixing matrix\cite{ck}.  If that is the case, one expects** to have a wealth of new CP violating phenomena in the next generation collider experiments at LHC or NLC.  This will be investigated in detail in the future.  Another interesting problem to be looked into more carefully in the future is the new CP violating phenomena predicted in this model.
 %%%%%%%%%%%%%%%%%%%%%%%%%%%%%%%%%%%%%%%%%%%%%%%%%%%%%%%%%%%%%%%%%%%%%%%%%%%%%%

\vskip 0.1in
{\it Acknowledgments}:
This research was supported by grants from National Science Council of Taiwan, ROC.  CHC also wish to thank National Center for Theoretical Sciences(NCTS), Physics Division at Hsinchu, Taiwan ROC and  Institute of Physics, Academia Sinica at Taipei, Taiwan ROC for partial support and hospitality.

\end{document}